\documentclass[a4paper,pra,showpacs,superscriptaddress,twocolumn]{revtex4-1}
\usepackage{hyperref}
\usepackage{amsfonts,dsfont,mathrsfs,amsmath,amssymb,bm,bbm}
\usepackage[english]{babel}
\usepackage[babel]{microtype}  
\usepackage{enumerate}
\usepackage{float}
\usepackage{subfigure}
\usepackage{braket}
\usepackage{graphicx}
\usepackage{svg}
\usepackage[ruled]{algorithm2e}
\usepackage{hyperref}

\newcommand{\myfig}[1]{Fig.\hspace{0.3em}\ref{#1}}

\begin{document}

	\title{Quantum Sparse Coding and Decoding Based on Quantum Network}
	
	\author{Xun Ji}
	\affiliation{National Laboratory of Solid State Microstructure, School of Physics, and Collaborative Innovation Center of Advanced Microstructures, Nanjing University, Nanjing 210093, China}
	\affiliation{Institute for Brain Sciences and Kuang Yaming Honors School, Nanjing University, Nanjing 210023, China}
    \affiliation{Hefei National Laboratory, Hefei 230088, China}
    
	\author{Qin Liu}
	\affiliation{National Laboratory of Solid State Microstructure, School of Physics, and Collaborative Innovation Center of Advanced Microstructures, Nanjing University, Nanjing 210093, China}
	\affiliation{Institute for Brain Sciences and Kuang Yaming Honors School, Nanjing University, Nanjing 210023, China}
    \affiliation{Hefei National Laboratory, Hefei 230088, China}
	
	\author{Shan Huang}
	\affiliation{National Laboratory of Solid State Microstructure, School of Physics, and Collaborative Innovation Center of Advanced Microstructures, Nanjing University, Nanjing 210093, China}
	\affiliation{Institute for Brain Sciences and Kuang Yaming Honors School, Nanjing University, Nanjing 210023, China}
    \affiliation{Hefei National Laboratory, Hefei 230088, China}

	\author{Andi Chen}
	\affiliation{National Laboratory of Solid State Microstructure, School of Physics, and Collaborative Innovation Center of Advanced Microstructures, Nanjing University, Nanjing 210093, China}
	\affiliation{Institute for Brain Sciences and Kuang Yaming Honors School, Nanjing University, Nanjing 210023, China}
    \affiliation{Hefei National Laboratory, Hefei 230088, China}

	\author{Shengjun Wu}
	\email{sjwu@nju.edu.cn}
	\affiliation{National Laboratory of Solid State Microstructure, School of Physics, and Collaborative Innovation Center of Advanced Microstructures, Nanjing University, Nanjing 210093, China}
	\affiliation{Institute for Brain Sciences and Kuang Yaming Honors School, Nanjing University, Nanjing 210023, China}
    \affiliation{Hefei National Laboratory, Hefei 230088, China}

    \begin{abstract}
    Sparse coding provides a versatile framework for efficiently capturing and representing crucial data (information) concisely, which plays an essential role in various computer science fields, including data compression, feature extraction, and general signal processing. In this study, we propose a symmetric quantum neural network for realizing sparse coding and decoding algorithms. Our networks consist of multi-layer, two-level unitary transformations that are naturally suited for optical circuits. Each gate is described by two real parameters, corresponding to reflectivity and phase shift. Specifically, the two networks can be efficiently trained together or separately using a quantum natural gradient descent algorithm, either simultaneously or independently. Utilizing the trained model, we achieve sparse coding and decoding of binary and grayscale images in classical problems, as well as that of complex quantum states in quantum problems separately. The results demonstrate an accuracy of 98.77\% for image reconstruction and a fidelity of 97.68\% for quantum state revivification. Our quantum sparse coding and decoding model offers improved generalization and robustness compared to the classical model, laying the groundwork for widespread practical applications in the emerging quantum era.
    
    \end{abstract}
	\maketitle
 
	\noindent \textbf{Introduction}\\
     In the classical field of machine learning and data analysis algorithms, the corresponding machine learning task is accomplished by combining supervised or unsupervised training on data \cite{WOS:000355286600030}. In quantum machine learning, various quantum versions of machine learning algorithms have also been proposed and developed immediately, including quantum support vector machines, quantum principal component analysis, quantum neural networks (QNN), and other quantum machine learning algorithms \cite{WOS:000410555900032,WOS:000341820700013,WOS:000500574300022}. Additionally, for rapidly evolving quantum neural network algorithms, classical optimizers are utilized to train multi-layer parameterized quantum gates in order to achieve desired learning tasks \cite{WOS:000993620700002}. Therefore, combined with Classical Sparse Coding (CSC) and quantum computation theory, an efficient quantum algorithm based on QNN can be designed to realize quantum sparse coding and decoding \cite{WOS:000493128300001,WOS:A1995QT40100017,WOS:000261782000008,bellante2022quantum}.

	Sparse coding and decoding refers to compressing high-dimensional data sets into their lower-dimensional representations sparsely and reconstructing higher-dimensional data sets from low-dimensional ones \cite{WOS:000961147200001,WOS:000406868800017,WOS:000258641500006}, which plays a crucial role in the fields of image compression, storage and processing \cite{WOS:000085444700052,WOS:000486237600006,WOS:000277884900014}. Sparse coding originates from the fact that the receptive field of the visual stripe cortex produces a sparse response to visual information. To be specific, most neurons of the visual stripe cortex are in a static resting state, while a small number of neurons are in the stimulated state \cite{WOS:A1996UQ65700051}. Mathematically, based on data structure analysis, digital image processing, and advanced algebra, CSC algorithms are developed gradually, where an overcomplete dictionary set is trained by various training algorithms, and sparse representation is a linear combination of the overcomplete dictionary set and active neurons in a high-dimensional space originally with redundant information \cite{WOS:000277884900014}. More explicitly, the sparse solutions of underdetermined equations are able to compress the data to a certain extent allowing little information loss of restoration. Currently, CSC algorithms have been implemented by hardware friendly, owing to their simple linear matrix operation in classical computer. 
 
	In contrast, quantum computing is an emerging frontier with higher information security and faster parallel acceleration. Specifically, quantum algorithms are expected for its potential advantages in quantum transmission, storage and encryption \cite{WOS:000343620500001,WOS:000431301800015,WOS:000454369600010,WOS:000514356400004,WOS:000209372300002,WOS:000340140200011,WOS:000646064000002,WOS:000888558600009,WOS:000685625800001,WOS:000935994100004}. Therefore, with the breakthrough of quantum machine learning algorithms and quantum circuits \cite{WOS:000410555900033,WOS:000390793900027,WOS:000476555600001}, more cutting-edge quantum algorithms are gradually moving toward hardware \cite{WOS:000940796200001,WOS:000711864600001,WOS:000367200400001,WOS:000340611600001,WOS:A1994NV63100015,WOS:000830702800001,WOS:000604141000011,WOS:000471037100002,WOS:000491132700075,WOS:000467374300005,WOS:000466034400001,WOS:000457547900068}, which is currently a possible development that challenges Moore's Law \cite{WOS:000888210300020}. Meanwhile, there is no efficient CSC algorithm to handle quantum states. Inspired by CSC algorithms \cite{WOS:000486237600006} and quantum circuit models \cite{WOS:000390793900027,WOS:A1995QT40100017}, we propose a quantum algorithm named Quantum Sparse Coding and Decoding (QSCD), which can sparsely code quantum states from high to low dimensional Hilbert space and then decode them inversely. Specifically, the quantum networks are trained by gradient-based descent method \cite{WOS:000444202100003,WOS:000246314600001,WOS:001001538500030}, and can be realized by universal multi-port design interferometers in optical quantum circuits \cite{WOS:000390793900027,WOS:000316614600026,WOS:000289982600014}. The simulations fully demonstrate that our QSCD algorithm is able to deal with the real and complex quantum states efficiently, which can be applied to further industrial problems as a cutting-edge technology.\\
	
	\begin{figure*}[ht]
		\centering
		\includegraphics[scale=0.615]{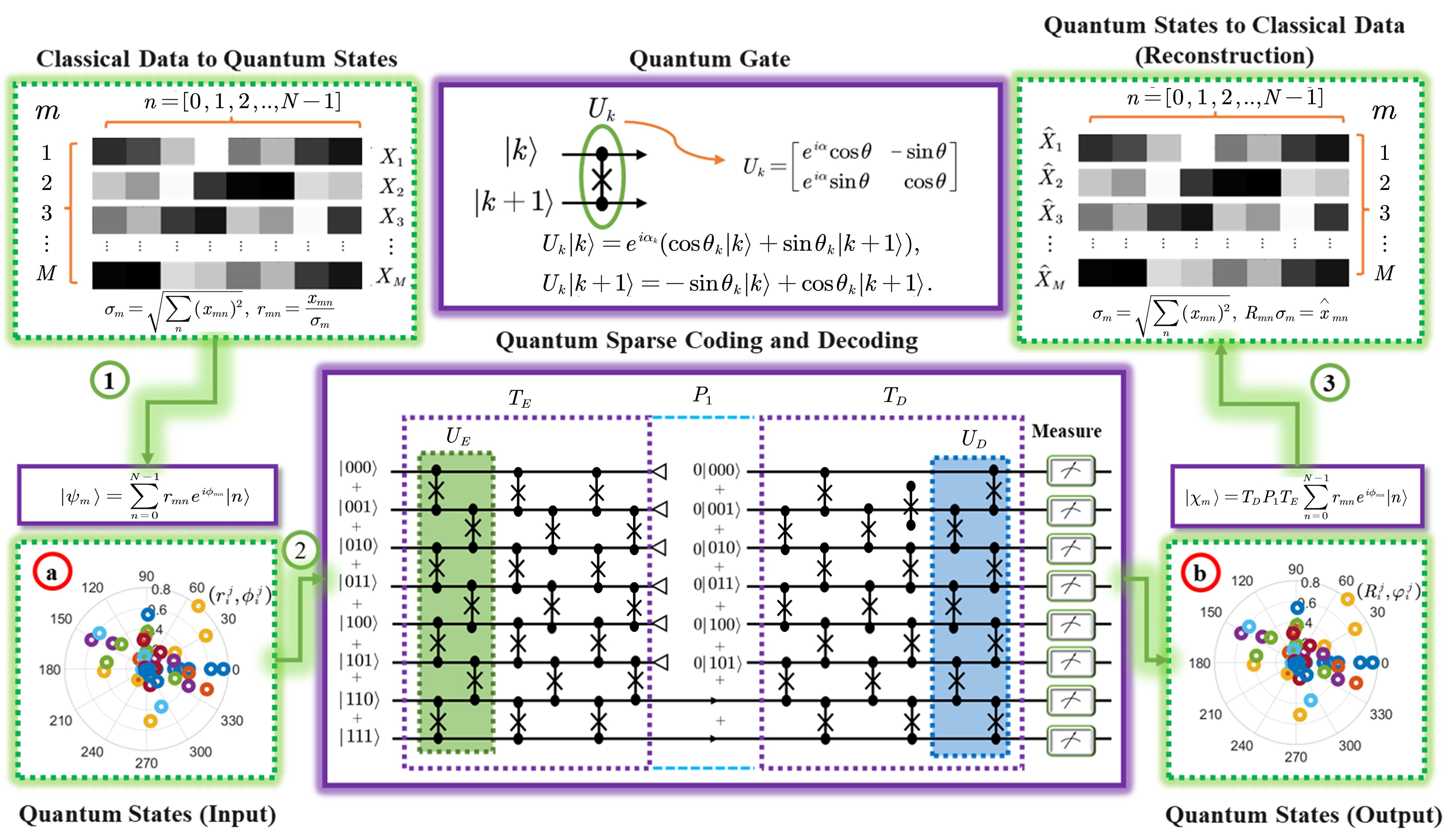}
		\caption{\textbf{The QSCD algorithm for grayscale image reconstruction.} In the one iteration, for image data, \textcircled{\scriptsize{1}} the classical information in a binary or grayscale image is converted into a column vector $X_m$ with $X_m \in[0,1]$, encoded into amplitude $r_{mn} \in[0,1]$ and phase $\phi_{mn} \in[0,2 \pi]$ of the real quantum state with $\phi_{mn}=0$. \textcircled{\scriptsize{2}}  The coded quantum state $\left|\psi_m\right\rangle$ is input into the coding and decoding network. \textcircled{\scriptsize{3}}  Then according to the measurement of the output real state, the amplitude is inversely decoded into the classical information $\hat{X}_m$ with reconstruction image pixel $\hat{x}_{mn}$. Similarly, for complex quantum states $\sum_{n=0}^{N-1} r_{mn} e^{i \phi_{mn}}|n\rangle$ in polar coordinates, corresponding to amplitude $r_{mn}$ and phase $\phi_{mn}$ in (\textbf{a}), it also can be directly input into the network. During the training process, (\textbf{b}) the probability amplitude $\left|R_{mn}\right|^2$ can be measured, which is shown in the polar coordinate form in the framework. Surely, for the trained quantum network, the sparse coding and decoding of the quantum states can be achieved without immediate measurement.}
		\label{F1}
	\end{figure*}

	\noindent \textbf{Results}\\
	\noindent \textbf{Preliminary.} 
    In classical information processing, data bits are encoded into binary electrical signals such as the ``high" and ``low" values of electric potential, facilitating convenient sparse representation and retrieval based on electrical circuits. As for the quantum information processing, the quantum counterpart of bits are two-dimensional quantum states, that is, quantum bits (qubits). Physical realizations of qubits include two-level atoms,  two distinct optical modes, etc. Taking the grayscale image reconstruction as an example, our proposal of QSCD is illustrated in \myfig{F1}. Consider $M$ grayscale images consisting of  $N$ pixels, and let $x_{mn}\in [0,1]$ denote the $n$th pixel of the $m$th grayscale image. From each grayscale image one then obtains a normalized $N$-dimensional row vector $\left\{r_{mn}\right\}$, where,  with $\sigma_m=\sqrt{\sum_n(x_{mn})^2}$,
 
    \begin{equation}\label{sigma}
        r_{mn}=\frac{x_{mn}}{\sigma_m}
    \end{equation}
are the normalized pixels which satisfy $\sum_{n=0}^{N-1} r_{mn}^2=1$ for all $m=1,\cdots,M$. It is worth noting that $\left\{r_{mn}\right\}$ alone does not contain the complete information of the original pixels, but when combined with the normalization factor $\sigma_m$, all the pixels of the $m$th grayscale image can be fully reconstructed.

    
In quantum sparse coding of grayscale images, we only need to focus on the normalized pixel vector $\left\{r_{mn}\right\}$ as it contributes to almost all the cost associated with image transmission and storage. It is convenient to write a general $N$-dimensional quantum state in the computational basis as follows
	\begin{align}\label{eq1}
		\left|\psi\right\rangle=\sum_{n=0}^{N-1} r_{n} e^{i \phi_{n}}|n\rangle.
	\end{align}
Here, each basis state $|n\rangle \in\{|00 \cdots 00\rangle,\ |00 \cdots 01\rangle,\ \ldots,\\\ |11 \cdots 11\rangle\}$ corresponds to the binary representation of an integer $n\in[0,N-1]$, and the expansion coefficients are normalized, $\sum_nr_n^2=1$.  In this way, classical information can be properly encoded into the expansion coefficients $\{r_ne^{i\phi_n}\}$ of $\ket{\psi}$ which, further serves as an input of our quantum sparse coding network. In particular, the quantum representation of normalized pixel vectors $\{\psi_m\}$ now become $\ket{\psi_m}=\sum_nr_{mn} e^{i \phi_{mn}}|n\rangle$, with phase parameters being 0, $\phi_{mn}=0$. 

     
    We proceed to detail our network architecture. Our sparse coding network aims to transform a collection of $N$-dimensional quantum states into lower dimensional ones while preserving, as much as possible, the information encoded in the input states. Considering that distinct classical information must be encoded in orthogonal states to be perfectly distinguishable under quantum measurements, our basic assumption in this work is that a coding network preserves orthogonality. In other words, we focus on quantum networks that implement unitary transformations. Furthermore, our main interest is in two particular types of quantum network structures, called Order type and Cross type. Order type links quantum gates sequentially, while Cross type interconnects them in a cross-pattern.  Numerical analyses show Order type excels at lower layers, whereas Cross type is better for deeper networks (see supplementary information I). 


	General unitary transformations on an $N$-dimensional Hilbert space $\mathcal{H}_N$ can be factored into a series of unitaries on 2-dimensional subspaces of $\mathcal{H}_N$ (single-qubit gates), which leave quantum states in the respective $(N-2)$-dimensional complementary subspaces invariant. Each specific factoring then determines a specific network structure, and the network parameters are exactly the variables describing individual single-qubit unitary gates. Notably, the network structures based on different ways of factoring unitary transformations may have different advantages and limitations.  In the following, our discussions will revolve around the optimal factoring proposed in Cross-type network (see supplementary information I), which is comparatively more efficient for deep networks. Its physical realization based on optical circuits consists of  $l_E$ layers of unitary gates on two neighboring optical modes, involving approximately $(N-1)l_E$  unitary gates in total \cite{WOS:000390793900027}. 
More specifically, a unitary gate $U_n$ on two optical modes $\{\ket{n},\ \ket{n+1}\}$ can be implemented by a phase shifter at one mode (with phase parameter $\alpha_n\in[0,2\pi]$) followed by a two-mode Mach-Zehnder interferometer (with parameter $\theta_n\in[0,\pi/2]$)
\begin{align}
    &U_n\ket{n}= e^{i\alpha_n}(\cos\theta_n|n\rangle+\sin\theta_n|n+1\rangle), \nonumber\\
    &U_n\ket{n+1}= -\sin\theta_n|n\rangle+\cos\theta_n|n+1\rangle.
\end{align}

Each layer of our coding network consists of $N/2$ parallel single-qubit unitary gates on the optical modes $\{\ket{2(k-1)},\ket{2k-1}\}\ (k=1,2,\cdots,N/2)$, followed by another $N/2-1$ parallel single-qubit unitary gates on the modes $\{\ket{2k-1)},\ket{2k}\}\ (k=1,2,\cdots,N/2-1)$ (see \myfig{F1} for an example and supplementary information I for more details). Then, the $l$-th layer of our coding network implements the transformation
	\begin{align} \label{eq4}
		U_E^{(l)}=&U_{0}U_{2}\cdots U_{N-2}U_1U_3\cdots U_{N-3} \nonumber\\
                =&\prod_{k=1}^{N-1} U_{k}\left(\theta_{k}^l, \alpha_{k}^l\right),
  \end{align}
 The overall coding network then essentially implements $l_E$ unitary transformations consecutively
	\begin{align}\label{eq3}
		T_{E}=U_E^{(l_{E})}U_E^{(l_E-1)}\cdots U_E^{(1)}:=\prod_{l=1}^{l_E} U_E^{(l)},
	\end{align}
 which is described by $2 l_E \times(N-1)$ real parameters \cite{WOS:000390793900027}. In supplementary I, we also discuss another network structure and compare it with the above one numerically.

	As for the decoding network, a direct choice is simply the mirror inversion of the coding network, which performs the inverse transformation to recover the initial states from the output of the coding network. 
 However, if some input state \eqref{eq1} is not transformed into the output space completely, say a $d$-dimensional Hilbert space $\mathcal{H}_d$  spanned by the last $d$ optical modes, then the actual outputs are described by the (unnormalized) states 
    \begin{align} \label{eq5}
    	\ket{\chi_m}=P_0T_E|\psi _m\rangle =P_0T_E\sum_{n=0}^{N-1} r_{mn} e^{i \phi_{mn}} |n\rangle,
    \end{align}
    with $P_0=\sum_{n=0}^{d-1}|n\rangle\langle n|$ the projector onto $\mathcal{H}_d$.  This projector is critical for realizing sparse coding. It retains prominent features of the initial set of states while discarding unimportant information.
 In accordance,  input photons of the decoding network are prepared into the normalized states $\{\braket{\chi_m|\chi_m}^{-\frac{1}{2}}\ket{\chi_m}\}$ which, after some decoding transformation $T_D$, become
    \begin{equation} \label{eq7}
        \begin{split}
        		\left|\Psi_m\right\rangle=&\frac{1}{\sqrt{\braket{\chi_m|\chi_m}}}T_D\ket{\chi_m}=\sum_{n=0}^{N-1} R_{mn} e^{i\varphi_{mn}}|n\rangle.
          \end{split}
    \end{equation}
In the above, $\{R_{mn}\}$ are real parameters, and $R_{mn}^2$ is the probability of detecting a photon at the $n$th output mode when prepared in the $m$th initial state.

We introduce the overall loss function $L$ for reconstructing the initial states $\{\ket{\psi_m}\}$  from outputs of the decoding network as follows
	\begin{align} \label{eq8}
		L=\sum_{m=1}^M\sum_{n=0}^{N-1} \left(JJ^*\right),
	\end{align}
where $J=R_{mn} e^{i \varphi_{mn}}-r_{mn} e^{i \phi_{mn}}$. With this loss function, we can update the parameters in our network during the training process to minimize potential errors. In the ideal case $L=0$, the information encoded in the initial states can be fully extracted without error. If the classical information is encoded only in the real coefficients $\{r_{mn}\}$ of the initial states $\{\ket{\psi_m}\}$  \eqref{eq1} and the decoding network implements the inverse transformation, a simpler and more convenient loss function is 
\begin{align}
    L_{inv}=1-\braket{\chi_m|\chi_m}=\braket{\psi_m|T_E^{-1}P_1T_E|\psi_m},
\end{align}
 with $P_1=\sum_{n=d}^{N-1}|n\rangle\langle n|$ denoting the projection onto the subspace of $\mathcal{H}_N$ that is complementary to the output space $\mathcal{H}_d$. Note that $P_0$ and $P_1$ sum up to the identity operator on $\mathcal{H}_N$ and quantum states are normalized  $\braket{\psi_m|T_E^{-1}P_0T_E|\psi_m}+\braket{\psi_m|T_E^{-1}P_1T_E|\psi_m}=1$, thus $L_{inv}$ is exactly the probability that output photons are not detected at the last $d$ modes.  $L_{inv}=0$ necessarily means that the encoding network successfully transforms all the initial states into the output space $\mathcal{H}_d$, $\ket{\chi_m}=T_E\ket{\psi_m}$, and the decoding network would perfectly recover those initial states.

     We remark here that if the initial states span a space with a dimension larger than the dimension $d$ of the output space, the loss function $L_{inv}$ would never approach  0 in the training process. Consequently, there is also no reason to believe that a decoding network implementing the inverse transformation $T_D=T_E^{-1}$ is optimal for minimizing $L_{inv}$ or, equivalently, maximizing the fidelity between the initial states and final states. This is in fact a much more general situation.  To mitigate error, we may properly train the decoding network to exploit, besides the features of individual input states,  also the structure of the input set. Moreover,  properly extending the depths of our coding and decoding networks is expected to further enhance the overall fidelity.

 
	In conclusion, we fashion the decoding network of the QSCD algorithm either by inverting the coding network or through parameter training of the decoding network. In the case of the decoding network with parameter training, the QSCD algorithm can be smoothly implemented in experiments, ensuring that parameter values remain within their designated range. Additionally, our approach to updating network parameters allows for concurrent updating of both the sparse coding network and the decoding network, solely through measuring the output states of the decoding network, eliminating the need for intermediate state measurements.\\

\begin{figure*}[ht]
		\centering
		\includegraphics[scale=0.159]{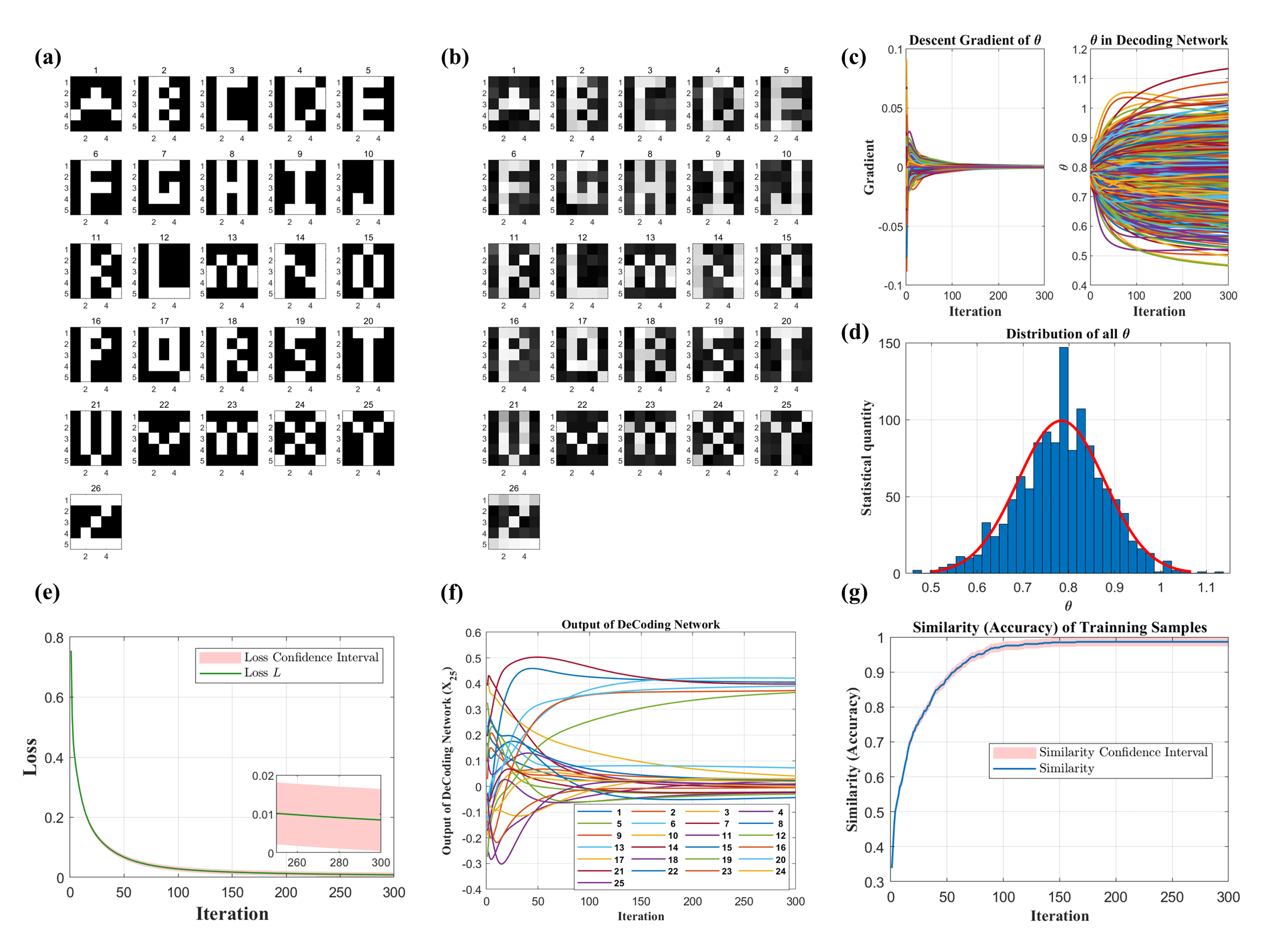}
		\caption{\textbf{Image reconstruction based on QSCD algorithm.} $\mathbf{a}$ The input image with binary value in $\{0,1\}$. $\mathbf{b}$ The output image with grayscale value in the range of $[0,1]$. During actually measuring the output quantum state in each iteration, the output of each iteration is always a real number in the range of $[0,1]$ owing to the floating parameters. Therefore, the reconstruction results of binary images are gray images, which illustrates the training adaptability of the network. $\mathbf{c}$ The update gradients of $\theta$ in 300 training iterations. At about the 100th training iteration, the quantum gate parameters tend to be stable with the gradients almost 0. $\mathbf{d}$ The distribution of optimal $\theta$. This distribution conforms to a normal distribution. $\mathbf{e}$ The loss of training network with $N \times M$-dimensional matrix $X$. The minimum MSE loss $L$ is 0.00873. $\mathbf{f}$ The training process of the sample letter "Y" $\left(X_{25}\right)$. The probability amplitude $R_{25}$ eventually tends to be in the range of $[-1,1]$. $\mathbf{g}$ The similarity of overall samples ("A"-"Z") between input images and output images in each iteration. After 300 training iterations that run for $895.234 \mathrm{~s}$ (CPU runs), the final similarity between input and output images is $98.77\%$.}
		\label{F2}
	\end{figure*}

	\noindent \textbf{Image Reconstruction.}
	For sparse coding and decoding of classical image information, the QSCD can be realized more efficiently. As shown in \myfig{F2}, firstly, the $D_1 \times D_2$-dimensional binary or grayscale image is converted into a column vector $X_m$ with $N \times 1$ dimension $\left(N \geq D_1 D_2>N / 2,N=2^q, q \in \mathbb{N}^+\right)$. Further, $X_m$ is encoded to a $N$-dimensional real quantum state $\left(\forall j \in\left[D_1 D_2, N-1\right], x_{mn} \equiv 0\right)$, so $M$ pictures are encoded to be the $N \times M$-dimensional matrix $X$. With $\sum_{n=0}^{N-1}\left(r_{mn}\right)^2=1$ and setting all $\phi_{mn}=0$, the encoded quantum state can be represented as $\left|\psi_m\right\rangle=\sum_{n=0}^{N-1} r_{mn}|n\rangle$ (see supplementary information IV and V). The probability amplitude is equal to the normalized value of the image grayscale value corresponding to Eq. \eqref{sigma}. The code $r_{mn}$ can be obtained to facilitate the preparation of quantum states.
	
    To see the performance of our network numerically, next we apply it to the $5 \times 5$ binary images of the 26 English capital letters as shown in  \myfig{F2}a. Each binary image requires 5 qubits to be faithfully represented $\left(D_1 D_2=25, N=2^5\right)$. 
    We choose a coding network with depth $l_E=20$ and a decoding network with depth $l_D=25$, the number of sparse coding channels is $d=4$. Thus, the coding network and decoding network involve $480$ and $600$ rotation parameters $\theta_k^l$, respectively (see supplementary information II). 
    Further,  we set the learning rate to be $\eta=0.01$ and update the network parameters through 150 iterations. 
    In each iteration, the output states of the decoding network are described by Eq. \eqref{eq7}, with which we are able to compute the respective loss function according to Eq. \eqref{eq8}. After each iteration, we then utilize the Quantum Natural Gradient Descent (QNGD) algorithm (see supplementary information III) to update the network parameters to reduce loss.
	The evolution of the real quantum states can be implemented by optical circuits to train and adjust the quantum gate parameters (see Methods). On the basis of the decrease of the loss function (\myfig{F2}c.), the reconstruction loss gradually tends to 0, and the gradient renewal of $\theta$ gradually approaches 0 (\myfig{F2}d), with the parameters $\theta$ gradually leveling off in the ranges of $\left[0, \frac{\pi}{2}\right]$. Eventually, the quantum network training is completed. The measurement results of output quantum states are converted to classical data $\hat{x}_{mn}$ (\myfig{F2}b):
	\begin{align}\label{eq12}
		\hat{x}_{mn}=\sigma_{m}R_{mn},
	\end{align}
    where $\sigma_m=(\sum_nx_{mn}^2)^{1/2}$ is normalization factor of the respective pixel vector $X_m$ as defined in Eq. \eqref{sigma}. The implementations mean that the classical information can be sparsely encoded and decoded through the QSCD algorithm, where the data flow is $X_m \rightarrow r_{mn} \rightarrow  \left|\psi_m\right\rangle \rightarrow T_D P_1 T_E\left|\psi_m\right\rangle \rightarrow R_{mn} \rightarrow \widehat{X}_m$. Undoubtedly, it also makes sense to achieve sparse coding and decoding of high-frequency or low-frequency signals, which provides effective ideas for pattern recognition.\\
	
 	\begin{figure*}[ht]
		\centering
		\includegraphics[scale=0.415]{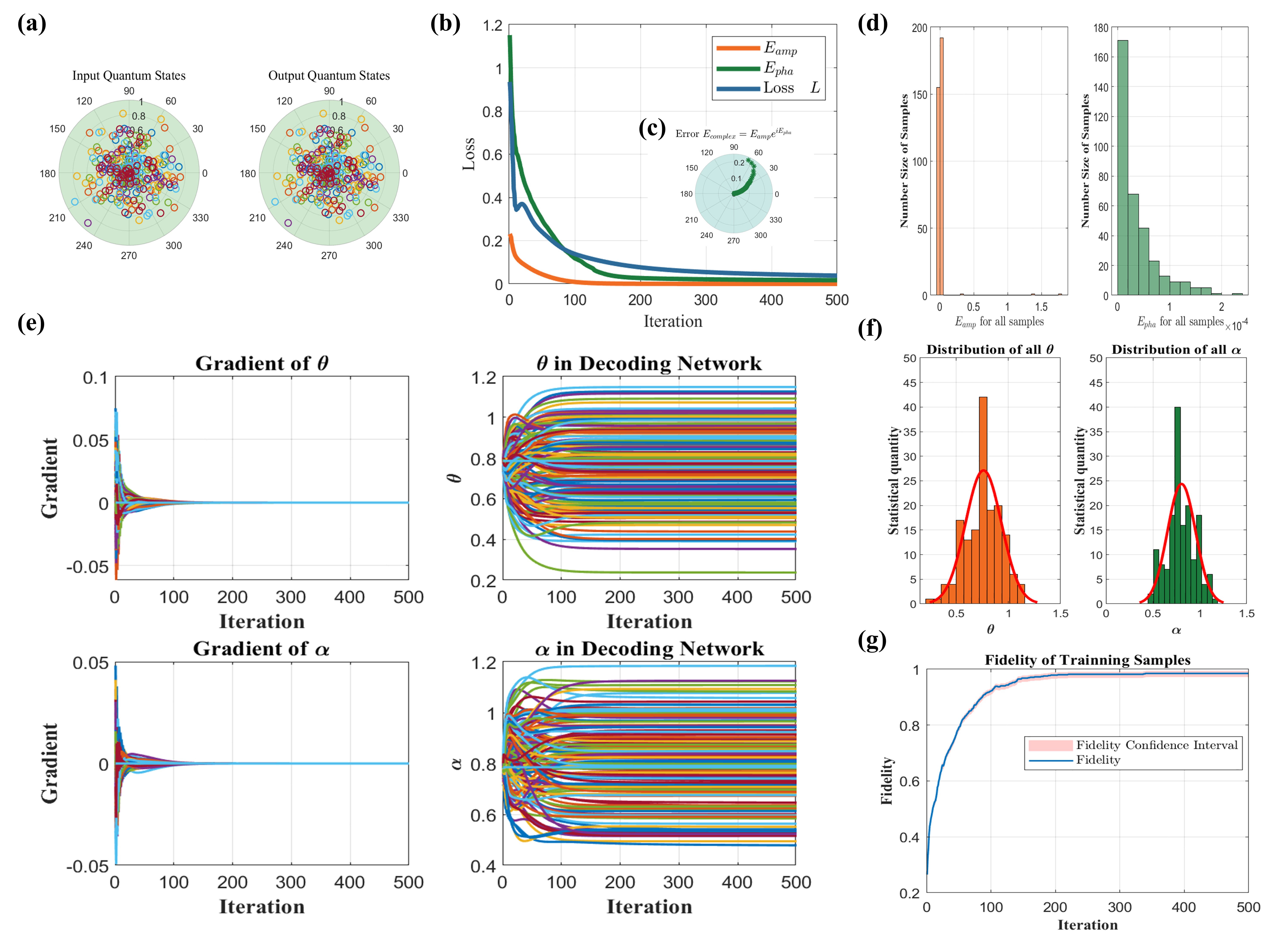}
		\caption{\textbf{Sparse coding and decoding for complex quantum states.} $\mathbf{a}$ The input and output quantum states. The amplitude and phase of the input and output quantum states are presented by measurement in polar coordinates. $\mathbf{b}$ The amplitude error $E_{amp}$ and phase error $E_{pha}$ in training iterations. The minimum amplitude error $\min E_{amp}=1.01 \times 10^{-4}$, The minimum phase error $\min E_{pha}=0.0143$. $\mathbf{c}$ The complex error in polar coordinates. The minimum complex error is $\min E_{ {complex}}=(0.99+i)10^{-4}$. $\mathbf{d}$ The amplitude and phase error of 50 quantum state samples. The amplitude and phase error of 50 quantum state samples are given by calculation. $\mathbf{e}$ ALL of $\theta$ and $\alpha$ and their gradients in the decoding network. $\mathbf{f}$ The distribution of optimal $\theta$ and $\alpha$. $\mathbf{g}$ The fidelity of training samples. The fidelity of the final quantum states is $97.68\%$.}
		\label{F3}
	\end{figure*}

	\noindent \textbf{Sparse Coding and Decoding for Complex Quantum States.} 
	For solving quantum problems, sparse coding and decoding of quantum states play a significant role in quantum cryptography. we achieve the simulation of sparse coding and decoding for complex quantum states in this section. Firstly, we generate 50 3-qubit quantum states in 8-dimensional Hilbert space ( $q=3, N=8, M=50$ ) as samples, and then they are input into the $T_E$ and $T_D$ (or $T_E^{-1}$ ) to train parameters. Set network parameters: $l_E=10, l_D=12, \eta=0.01, d=4$ and $Ite=500$. By measuring the output states of the decoding network in each training iteration, the loss is calculated, so as to update the quantum gate parameters in Eq. \eqref{eq7}. Obviously, we need to train both the phase shift $\alpha_k^l$ and reflectivity $\cos \theta_k^l$ (see supplementary information II, \myfig{F3}a). Therefore, the quantum network needs to be trained with $140$ parameters $\left(\theta_k^l, \alpha_k^l\right)$ of the coding network and $168$ parameters $\left(\hat{\theta}_k^l, \hat{\alpha}_k^l\right)$ of the decoding network.
	In the quantum sparse coding network $T_E$, only 4 output optical circuits are connected to the decoding network $T_D$. The amplitude error $E_{\rm amp}$ and phase error $E_{\rm pha}$ are defined as $E_{\rm amp}=$ $ \sum_{m=1}^M\sum_{n=0}^{N-1}\left(R_{mn}-r_{mn}\right)^2, E_{\rm pha}=\sum_{m=1}^M\sum_{n=0}^{N-1}\left(\varphi_{mn}-\phi_{mn}\right)^2$. As shown in \myfig{F3}b, \myfig{F3}c and \myfig{F3}d, the $R_{mn}$ and $\varphi_{mn}$ of the output quantum states are measured, and the error between output and input quantum states is obtained. Obviously, the initial phase error is significantly larger than the amplitude error (\myfig{F3}b), because for each quantum state in a quantum circuit, the probability amplitude $R_{mn}$ is much less than 1 (even most of which is in the range of $[0.1,0.2])$. Furthermore, $R_{mn}$ and $\varphi_{mn}$ error of the final states will gradually range to 0 during continuous training. Finally, after 500 training iterations, the output states are similar to the input states in amplitude and phase. Another more intuitive error in polar coordinates in \myfig{F3}c, represents the complex error $\left(E_{ {\rm complex }}=E_{\rm amp} e^{iE_{\rm pha}}\right)$, which gradually tends to 0 with the slope around 0 in \myfig{F3}c.After training, the optimal results are obtained, in which input states and output states are similar in \myfig{F3}a. Therefore, the error of the ultimate reconstruction will be calculated, and then $E_{\rm amp}$
	and $E_{\rm pha}$ of the overall samples are extremely close to 0 in \myfig{F3}d.
	For the QSCD algorithm, only $R_{mn}$ and $\varphi_{mn}$ of the output quantum states are measured. Therefore, sparse coding and decoding of quantum states are able to be implemented in integrated optical circuits, simutaneously the elimination of the intermediate measurement can greatly reduce the system error. Using QSCD to achieve sparse coding and decoding of quantum states guarantees the transmission efficiency of quantum information.\\

	\noindent \textbf{Discussion}\\
	Generally, the QSCD algorithm can be implemented in physical hardware efficiently.  It can achieve sparse coding and decoding by encoding classical image data (derived from the pixels of binary and grayscale images) into the amplitude of real quantum states. Furthermore, complex quantum states can also be sparsely coded and decoded by the QSCD algorithm through quantum unitary transformation and adaptive quantum natural gradient descent algorithm.
	
	The superiority of the QSCD algorithm over traditional QNN based on Hamiltonian evolution is embodied in its lower computation complexity, better hardware-friendly performance, and ability to deal with higher dimensional information. The current optical quantum computing devices are more suitable for sparse quantum unitary transformation in our QSCD network, which increases the experimental efficiency and is expected to solve more quantum problems such as calculating eigenstates. Moreover, the differences between QSCD and CSC root in their computation frameworks, where the QSCD is constructed by multiple sparse matrices with the same dimensions, and the CSC is composed of one matrix full of weight parameters. In more detail, the two real parameters are updated together in each sparse matrix in the QSCD algorithm, nevertheless, the full weights are adjusted in a dictionary set in the CSC algorithm.\\

	\noindent \textbf{Methods}\\
	\noindent \textbf{Quantum Gate Design.}
	In this paper, the QSCD algorithm is realized through quantum optical circuits. Specifically, a universal multi-port interferometer can be programmed to achieve any linear transformation between two channels by implementing a quantum gate $U_{(k, k+1)}$ in port $k$ and $k+1$, abbreviated to $U_k$. These interferometers typically consist of a regular grid of beam splitters and phase shifters and are manufactured directly with an integrated photon architecture and off-the-shelf scalability. At present, the standard design of universal multiport interferometers is based on the decomposing from any $N \times N$ unitary matrix to a linear combination of two-dimensional beam-splitting transform sequence \cite{WOS:000390793900027}. Compared with the traditional design, the main advantage of this design is only half the optical depth and more anti-optical loss (see supplementary information I).\\

	\begin{algorithm}[htbp] 
		\caption{Calculate the Quantum Sparse Coding and Decoding with respect to the $X$.}
        \label{A1}
		\begin{align*}
			&{\bf Input:}\ {X, l_E, l_{D} \text { to get }[M, N] \leftarrow \operatorname{size}(X)}\\
			&{\bf Output:}\ {\left(\theta_k^{l_E}, \alpha_k^{l_E}\right),\left(\theta_k^{l_{D}}, \alpha_k^{l_{D}}\right), g_E, g_{D}, L, \widehat{X}_i}\\
			&{\bf Initialize}\ (\theta,\alpha)\leftarrow (\pi / 3,2 \pi / 3),\text {Ite}=150, \eta=0.01, d=4\\
			&\textbf{for} \ {m=1:M}\\
			&\left|\psi_m\right\rangle \leftarrow X_m\\
			&\textbf{end}\\
			&\left[\Big(\theta_k^{l_E}, \alpha_k^{l_E}\Big),\left(\theta_k^{l_{D}}, \alpha_k^{l_{D}}\right), g_E, g_{D}, L\right]\\
			&=\operatorname{trainQSCD}\left(l_E, l_{D},|\psi\rangle,(\theta, \alpha),  {Ite}, \eta, d\right)\\
			&\textbf{For}\ {m=1:M}\\
			&\left|\hat{\Psi}_m\right\rangle=T_{D}\left(\theta_k^{l_{D}}, \alpha_k^{l_{D}}\right) P_1 T_E\left(\theta_k^{l_E}, \alpha_k^{l_E}\right)\left|\psi_m\right\rangle\\
			&\widehat{X}_m \leftarrow\left|\hat{\Psi}_m\right\rangle\\
			&{\bf end}
		\end{align*}
	\end{algorithm}

    \begin{algorithm}[htbp] 
		\caption{Calculate the loss function and the gradients to update $(\theta,\alpha)$, trainQSCD.}
        \label{A2}
		\begin{align*}
			&{\bf Input:}\ l_E,l_D,|\psi \rangle, \left( \theta ,\alpha \right) ,Ite,\eta ,d \text { to get }[M, N]\\ 
			&{\bf Output:}\ \left( \theta _{k}^{l_E},\alpha _{k}^{l_E} \right) ,\left( \theta _{k}^{l_D},\alpha _{k}^{l_D} \right) ,g_E,g_D,L\\
			&{\bf Initialize}\ \varDelta=10^{-8}\\ 
            &\left|\hat{\Psi}\right\rangle =T_D\left( \theta ,\alpha \right) P_1T_E\left( \theta ,\alpha \right) |\psi \rangle \\
            &\textbf{Update:}\ \left( \theta _{k}^{l_E},\alpha _{k}^{l_E} \right) \\
			&\textbf{for} \ {i=1:Ite}\\   
	&L=\left( R_{mn}-r_{mn} \right)^2 \\
            &\textbf{for} \ {l=1:l_E}\\
            &\textbf{for} \ {k=1:N}\\
            &\partial \theta _{k}^{l}  =\left[ T_DP_1T_E\left( \theta _{k}^{l}+\varDelta ,\alpha _{k}^{l} \right) -T_DP_1T_E\left( \theta _{k}^{l},\alpha _{k}^{l} \right) \right] /\varDelta \\
            &\partial \alpha _{k}^{l}  =\left[ T_DP_1T_E\left( \theta _{k}^{l},\alpha _{k}^{l}+\varDelta \right) -T_DP_1T_E\left( \theta _{k}^{l},\alpha _{k}^{l} \right) \right] /\varDelta \\
            &g_E\left( \theta_{k}^{l} \right)=sum\left( L\cdot \partial \left( \theta _{k}^{l} \right) \right) /\left( M\times N \right)\\
            &g_E\left( \alpha _{k}^{l} \right)=sum\left( L\cdot \partial \left( \alpha _{k}^{l} \right) \right) /\left( M\times N \right)\\
            &\theta _{k}^{l}=\theta _{k}^{l}-\eta g_E\left( \theta_{k}^{l}  \right) ,\alpha _{k}^{l}=\alpha _{k}^{l}-\eta g_E\left( \alpha_{k}^{l}  \right) \\
            &{\bf end}\\
            &{\bf end}\\
            &\textbf{same way to calculate}\ g_D\ \textbf{to update}\  \left( \theta _{k}^{l_D},\alpha _{k}^{l_D} \right)\ 
		\end{align*}
  
   \end{algorithm}
   
\vspace*{0.5cm}

	\begin{algorithm}[htbp] 
		\caption{Construct the quantum network $T_E$ and $T_D$.}%
        \label{A3}
		\begin{align*}
			&{\bf Input:}\ \left( \theta ,\alpha \right) ,[M,N]\hspace{10cm}\\%
			&{\bf Output:}\ T_E,T_D\\%
            &{\bf Initialize}\ U_k,U_E,T_E,U_E,T_E\\
			&\textbf{for} \ {l=1:M}\\
            &\theta\leftarrow\theta^l,\alpha\leftarrow\alpha^l\\
		  &\textbf{for} \ {k=1:N-1}\\
            &U_E^{\left( l \right)}=U_E^{\left( l \right)}U_k(\theta_k^l,\alpha_k^l)\\
			&{\bf end}\\
            &T_E=T_EU_E^{\left( l \right)}\\
            &\textbf{Reset:}\ U_E\\
            &{\bf end}\\
            &\textbf{same way to construct}\ U_D,T_D\ 
		\end{align*}
	\end{algorithm}

	\noindent \textbf{QSCD Algorithm.}
	According to the theory of the QSCD algorithm, the main program can be summarized as Algorithm. \ref{A1}. The main simulation programs are parameter training and quantum gate construction, which are simulated with Matlab programming language. As shown in Algorithm. \ref{A1}, for classical image data, there are data-to-state coding function and training function (trainQSCD). Given parameter initialization, the $r$ and $\phi$ of quantum states are input to train $(\theta, \alpha)$ of quantum network by Algorithm. \ref{A2}. After that, the program is presented in Algorithm. \ref{A3}. is called, then the output result is obtained by implementing the coding and decoding process $T_D P_1 T_E$. Finally, the program returns reconstruction results and procedure parameters.\\
    \\
 
	\noindent \textbf{Quantum Natural Gradient Descent Algorithm.}
	The $\operatorname{trainQSCD}$ pseudocode calculates the gradient of the $(\theta, \alpha)$, which is denoted as the gradient $g_E$ for network $T_E$ and $g_D$ for network $T_E$. In the process, the quantum natural gradient of quantum gate parameters $(\theta, \alpha)$ is calculated by the QNGD algorithm \cite{WOS:000940796200001}. Afterward, based on the update of parameter gradients, the renewal of parameters will be acted upon in the next training iteration. It is noticed that the parameters $(\theta, \alpha)$ are updated at the same time, but the adjustment of the reflectivity $\theta$ will not affect the adjustment of the phase $\alpha$ (see supplementary information II).\\
	
	\noindent \textbf{Quantum Network Construction.}
	Constructed with multi-layer rotation gate $U_k$, the network $T_E$ and $T_D$ can also be combined into a quantum network of any depth separately in Algorithm. \ref{A3}. The quantum network construction gives out the form of the unitary transformation, which can be in Order type and Cross type (see supplementary information I). In the network $T_D$, the transformation $U_D$ needs to be connected in reverse order of $U_E$ in sparse coding network $T_E$.\\

\newpage

    \noindent \textbf{CODE AVAILABILITY}\\
    Some of the simulation code can be found in 
    \href{https://github.com/Jixun97/QSCD.git}{https://github.com/Jixun97/QSCD.git}. The complete code used for simulations is available from the corresponding authors upon reasonable request. \\

    \noindent \textbf{ACKNOWLEDGMENTS}\\
    This work is supported by the National Natural Science Foundation of China (No. 12175104),  the National Key Research and Development Program of China (No. 2023YFC2205802), and the Innovation Program for Quantum Science and Technology (No. 2021ZD0301701).\\

    \noindent \textbf{AUTHOR CONTRIBUTIONS}\\
    X.J. and Q.L. and S.H. contributed equally. S.W. directed and supervised the project. S.W., X.J., and Q.L. proposed the QSCD algorithm and designed the implementations. X.J. performed the numerical simulation and analyzed the data with the assistance of Q.L., S.H., and A.C.. Q.L. and X.J. provided theoretical support under the guidance of S.W.. A.C. and Q.L. provided data support. X.J., Q.L., and S.H. wrote and revised the manuscript with feedback from all authors.\\
    
    \noindent \textbf{ADDITIONAL INFORMATION}\\
    \noindent \textbf{Supplementary information: }
    Supplementary Information to: Quantum Sparse Coding and Decoding Based on Quantum Network.\\
    \noindent \textbf{Supplementary training demo video: }A gray image training process is based on the QSCD algorithm.\\
    \noindent \textbf{Competing interests: }
    The authors declare no competing interests.

\newpage

\bibliographystyle{naturemag}

\bibliography{reference}

\end{document}